\documentclass[12pt]{iopart}
\usepackage{iopams}
\usepackage{epsfig}
\usepackage{epsf}
\def\be{\begin{equation}}
\def\ee{\end{equation}}
\def\bea{\begin{eqnarray}}
\def\eea{\end{eqnarray}}

\def\a{\alpha}
\def\b{\beta}
\def\d{\delta}
\def\e{\epsilon}

\def\k{\kappa}

\def\r{\rho}
\def\s{\sigma}

\def\nn{\noindent}

\begin{document}
\title{Soliton solutions of driven non-linear  and higher order non-linear Schr\"odinger equations}


\author{ Vivek M Vyas$^1$, T Soloman Raju
$^2$, C Nagaraja Kumar$^3$ and Prasanta K Panigrahi$^4$
\footnote[1]{\texttt{prasanta@prl.ernet.in}}}

\address{$^1$ Department of Physics, M S University of Baroda, Vadodara, 390 002, India}

\address{$^2$ Physics Group,
BITS-Pilani, Goa Campus, Zuari Nagar, Goa, 403 726, India}

\address{$^3$ Department of Physics,
Panjab University, Chandigarh, 160 014, India}

\address{$^4$Physical Research Laboratory, Navrangpura, Ahmedabad
380 009, India}

\begin{abstract}
We analyse the structure of the exact, dark and bright soliton
solutions of the driven  non-linear Schr\"odinger equation. It is
found that, a wide class of solutions of the higher order non-linear
Schr\"odinger equation with a source can also be obtained through
the above procedure. Distinct parameter ranges, allowing the
existence of these solutions, phase locked with their respective
sources, are delineated. Conditions for obtaining non-propagating
solutions are found to be quite different for both the equations. A
special case, where the scale of the soliton emerges as a free
parameter, is obtained and the condition under which solitons can
develop singularity is pointed out. We also study the highly
restrictive structure of the localised solutions, when the phase and
amplitude get coupled.

\end{abstract}




\section{Introduction}


The externally driven, non-linear Schr\"odinger equation (NLSE) with
a source has been investigated  in the context of a variety of
physical processes. It arises in the problem of Josephson junction
\cite{sam}, charge density waves \cite{newell}, twin-core optical
fibres \cite{snyder,boris,gil,solomon1}, plasma driven by rf fields
\cite{bekki1} and a number of other problems \cite{bekki2}. As
compared to NLSE, which is an integrable system \cite{das}, not much
is known about the exact solutions of this equation. Perturbative
solutions around the stable soliton solutions of NLSE with a source
have been studied earlier. Analysis around constant background and
numerical investigations \cite{smirnov,barashenkov,nistazakis} have
revealed the phenomenon of auto-resonance \cite{shagalov,friedland}
as a key characteristic of this system, where a continuous phase
locking between the solutions of NLSE and the driven field is
observed.

In a recent work \cite{pkp}, some of the present authors have
devised a procedure  based on fractional linear transformation for
obtaining exact solutions of this dynamical system. One obtains both
localised and oscillatory solutions. Apart from these regular
solutions, under certain constraints singular solutions have also
been found, implying extreme increase in the field intensity due to
self-focussing. This approach is non-perturbative in the sense that,
the obtained exact solutions are necessarily of rational type, with
both numerator and denominator containing terms quadratic in
elliptic functions. However, the exact parameter ranges in which the
general solution exhibits bright and dark nature has not been
investigated. Considering, the importance of the localised solutions
of this physically important dynamical system, the above point needs
a systematic study.

The goal of the present paper is to analyse in detail the structure
of the most general, localised bright and dark solitons of the NLSE
with a source. The parametric restrictions, under which singular
structures can form, are obtained.  Unlike NLSE, it is observed that
dark and bright solitons, depend both on coupling and source
strengths. For example, bright solitons can also form in the
repulsive regime, if the source strength has positive value.
Similarly dark solitons can form in the attractive regime.
Conditions which give non-propagating solutions, for driven NLSE,
are studied. The possibility, where the phase and amplitude can get
related is also investigated. The highly restrictive nature of the
resulting dynamics is pointed out. In the process, we also find a
large class of solutions to higher order nonlinear Schr\"odinger
equation (HNLSE) with a source, by connecting the solution space of
these two equations. HNLSE, which is a generalisation of NLSE, has
been proposed by Kodama \cite{kodama} and Kodama and Hasegawa
\cite{kh} to describe the propagation of short duration
(femtosecond) pulses in optical fibres. The fact that NLSE with a
source describes twin-core fibres, under certain conditions
\cite{gil,solomon1,solomon2}, motivates us to find exact solution of
HNLSE with a source. A possible scenario leading to driven HNLSE is
pulse propagation through asymmetric twin-core fibre involving a
femtosecond pulse in first core, having anomalous dispersion, and a
nanosecond pulse propagating in second core in the normal dispersion
regime. Analysis of the solution space of the driven HNLSE shows
many interesting features. Non-propagating solutions of driven HNLSE
show a much richer structure as compared to NLSE, because of the
presence of the higher order terms.


The paper is organised as follows. In the following section, we
elaborate briefly on the method of the fractional linear transform
for obtaining the solutions of the NLSE with a source. In this
section, it is shown that for a large class of solutions, driven
HNLSE can also be connected with driven NLSE. Hence, the obtained
solutions are also solutions of HNLSE, albeit with different
relations among the parameters. The third section is devoted to the
investigation of appropriate parameter regimes, characterising
bright, dark and singular solutions, and their analysis. In the
fourth section, the structure of the solutions space is studied,
when the phase and amplitudes are coupled. We then conclude after
pointing out the open problems and future directions of work.

\section{Analysis of NLSE phase locked with source\\
   and its relation to HNLSE}

The equation which we intent to solve is driven NLSE, which is space
time dependent, and phase locked with the source:
 {\be \label{nlse} i\frac{\partial \phi}{\partial t} +
\frac{\partial^{2} \phi}{\partial x^2} +g \mid \phi \mid^2 \phi +\mu
\phi=\kappa e^{i(kx - \omega t )}\quad , \ee} \nn where $g$, $\mu$,
$\kappa$, $\omega$ and $k$ are real constants. As we will see later,
it can be connected with HNLSE, for a wide class of solutions. We
consider the ansatz travelling wave solution in the form,

\bea \nonumber \label{ansatz} \phi(x,t) = \s(\xi) e^{i(k x -
\omega t)} ,\eea

\nn where $\xi =\alpha (x- v t)$. Separating the real and
imaginary parts of  equation (\ref{nlse}), one obtains, $ v=2k $,
from the imaginary part, indicating that in the present case, wave
velocity $v$ is controlled by $k$. The real part yields,

\be \label{realeq} \alpha^2 \s^{\prime\prime} +g\s^3+{\e{ \s}} =
{\kappa} ,\ee

\nn where $\epsilon=\omega-k^2+\mu$, and prime indicates
differentiation with respect to $\xi$. As has been observed
earlier \cite{pkp}, this equation can be connected to the equation
$f^{\prime\prime}+af+bf^3=0$ through the following fractional
linear transformation (FT):

\be \label{FT}   \s (\xi) =  \frac {A + B f(\xi|m)^{2}} {1+ D
f(\xi|m)^{2}}\ee

\nn where $A$,$B$ and $D$ are real constants, and
 $f(\xi|m)$ is a Jacobi elliptic function, with the modulus parameter m.
 We consider the case  where,
 $ f({\xi|m}) = cn({\xi}|m) $, other cases can be similarly studied.
 Since the goal is to study the localised solutions, we consider the case
 with modulus parameter $m = 1$, which
 reduces $cn(\xi)$ to sec hyperbolic of $\xi$. It is worth pointing out that,
  other solutions involving $sn(\xi)$ and
 $dn(\xi)$ naturally emerge from the above solution,
  since the transform involves square of the $cn(\xi)$
 function. It should also be mentioned that, in the above fractional transform the second
 power of the
 cnoidal functions emerge, as the highest power.

We can see that equation (\ref{FT}) connects $\s(\xi)$ to Jacobi
elliptic equation, provided $AD{\neq}B$, and the following
conditions are satisfied for the localised solutions:

 \bea \label{condi1}
 A\e +gA^{3}-\kappa=0,\\
\label{condi2}2\e AD+\e B+4\alpha^{2}(B-AD)+3gA^{2}B-3\kappa D=0,\\
\label{condi3}
 \nonumber A\e D^{2}+2\e BD+4\alpha^{2}(AD-B)D \\
  +6\a^{2}(AD-B)+3gAB^{2}-3\kappa D^{2}=0,\\
\label{condi4} \e B D^2+2\a^{2}(B-AD)D+gB^{3}-\kappa D^{3}=0.
 \eea

\nn  Equation (\ref{condi1}) in $A$ does not involve $B$ and $D$,
which is first solved to get the real $A$. Thus $A$ is determined in
terms of $\e$, $\kappa$, and $g$. From equation (\ref{condi2}), we
determine $D$ in terms of $B$ as,
 $D=\Gamma B,$ where,
$\Gamma=\frac{\e + 4\a^2 +3gA^2}{4\a^2 A+3\kappa -2\e A}$. By
substituting this in equation (\ref{condi3}), $B$ is found as
$B=\frac{6\a^2(1-A\Gamma)} {3gA+A\e\Gamma^2+2\e\Gamma+4\a^2
\Gamma(A\Gamma-1)-3\kappa\Gamma^2}$. From equation (\ref{condi4}),
we obtain a cubic equation in $\b\equiv\a^2$:
 \be \label{cubic1}
p_{1}\b^3+q_{1}\b^2+r_{1}\b+c_{1}=0, \ee
 where $p_1=64(A^3 g -A \e -\k)$, $q_1=(48 A^5 g^2 +64 A^3 g \e
 +16 A {\e}^2-48 A^2 g \k -16 \e \k)$,
$r_1=(12 A^7 g^3 + 36 A^5 g^2 \e +20 A^3 g {\e}^2 -4 A{\e}^3 -60
A^4 g^2 \k -72 A^2 g \e \k +4 {\e}^2 \k +48 A g {\k}^2 )$ and
$c_{1}=(3 A^7 g^3 \e -3 A^5 g^2 {\e}^2 -7 A^3 g {\e}^3 -A {\e}^4
-18 A^6 g^3 \k -15 A^4 g^2 \e \k +12 A^2 g {\e}^2 \k +{\e}^3 \k +9
A^3 g^2 {\k}^2 -15 A g \e {\k}^2 + 9 g {\k}^3) $. It can be
straightforwardly seen that $p_1$ in equation (\ref{cubic1}) is
the consistency condition (\ref{condi1}) and hence is identically
zero. Therefore, the width parameter $\b$ is the solution of a
quadratic equation. Thus for any given values of $g$, $\e$ and
$\kappa$, we can find the values of $A$, $B$, $D$ and $\alpha$.

Before continuing the analysis of the localised solutions, let us
turn to driven HNLSE:

\be \label{hnlse} i\frac{\partial \psi}{\partial t} +
\frac{\partial^{2} \psi}{\partial x^2} +g \mid \psi \mid^2 \psi +\mu
\psi + i \bar{\e} \left[ \tilde{A} \frac{\partial^{3} \psi}{\partial
x^3} +\tilde{B} \mid \psi\mid^2 \frac{\partial \psi}{\partial x}+
\tilde{C} \psi \frac{\partial{\mid \psi\mid^2}}{\partial x}\right]=
 \kappa e^{i(k x - \omega t )}, \ee
\nn where $g$, $\mu$, $\bar{\e}$, $\tilde{A}$, $\tilde{B}$,
$\tilde{C}$, $\omega$ and $k$ are real constants, and $\kappa$ is
the source strength.

The different nonlinear terms describe various interactions which
affect the propagation of femtosecond pulses in optical fibres
\cite{HKbook}. The term proportional to $\tilde{A}$ results from the
inclusion of the effect of third order dispersion and the term
proportional to $\tilde{B}$ comes from the first derivative of the
slowly varying part of nonlinear polarization, which is responsible
for self-steepening and shock formation. The delayed Raman response
for self-frequency shift accounts for the term proportional to
$\tilde{C}$.

In last many years lot of activity has gone into the study of the
solvability of the above equation without a source. Integrability
studies were performed using inverse scattering transform method and
Hirota method and a few cases of integrability has been
identified:(i) Sasa-Satsuma equation
($\tilde{A}$:$\tilde{B}$:$\tilde{C}$ $=$ $1$:$6$:$3$) \cite{S-S},
(ii) Hirota equation ($\tilde{A}$:$\tilde{B}$:$\tilde{C}$ $=$
$1$:$6$:$0$) \cite{hirota}, (iii) derivative NLSE, of type I and
type II \cite{A-C}. Several exact solutions of solitary wave (bright
soliton) and kink type (dark soliton) have been obtained
\cite{PT,palacios}. Phase modulated solutions to HNLSE have also
been obtained \cite{cnk}, where the generic form of the solution
reads $ \psi(x,t) = f (x,t) e^{i(\chi(x,t)- \omega t)} $; here
$\chi$ is some function of $x$ and $t$.
 Substituting the ansatz solution of the form,

 \be \label{ansatz} \psi(x,t) =  \rho(\xi) e^{i(k x -
\omega t)}.\ee

\nn in equation (\ref{hnlse}) and then separating out the imaginary
and real parts, we get the following two equations,

\be \label{imaghnlse}  {\a}^3 \e \tilde{A} \r''' + (2 \a k  -3 \a
\e k^2 \tilde{A}
 -\a v) \r'
  + (\a \e \tilde{B} + 2 \a \e \tilde{C}) \r^2 \r' = 0,\ee

 \be \label{realhnlse} {\tilde{\alpha}}^2
\rho^{\prime\prime} +\tilde{g}\rho^3+{\tilde{\epsilon}{ \rho}} =
{\kappa}.\ee

\nn Here the parameters are defined as,

\be \label{pr1} {\tilde{\alpha}}^2 = \a^2 ( 1 - {3 \bar{\e}
\tilde{A} k} ), \ee
 \be \label{pr2} \tilde{g}=(g -
{\bar{\e} \tilde{B} k }), \ee
\nn and
 \be
\label{pr3}\tilde{\epsilon} = (\omega - k^2 + \mu + \bar{\e}
\tilde{A} k^3). \ee It is interesting to note that, the above real
part of HNLSE is similar to the real part of driven NLSE.

 \nn Differentiating
equation (\ref{realhnlse}) once and comparing it with equation
(\ref{imaghnlse}), we see that they are consistent only if the
parameters satisfy the relations:

\be \label{concondi1} \tilde{B}+2\tilde{C}= \frac {3 {\alpha}^2
\tilde{A} g} {\tilde{\a}^2}\ee \nn and
 \be \label{concondi2} v = 2k - 3\bar{\e} \tilde{A} k^2 - \frac {{\alpha}^2
\bar{\e} \tilde{\e} \tilde{A}} {\tilde{{\alpha}^2}}. \ee

  Hence, for this class of solutions the problem
of solving equation (\ref{hnlse}) is mapped to the problem of
solving equation (\ref{realhnlse}), which is nothing but the real
part of the driven NLSE. This proves our assertion that phase locked
solutions of driven NLSE are mapped to the phase locked solutions of
driven HNLSE, in certain parameter ranges. One can see from the
above relation that the velocity in this case has a quadratic
relation with $k$, a feature very different from the driven NLSE.

\section{Analysis of the localised solutions}

We now proceed to analyse carefully the localised solutions of
equation (\ref{realeq}). It is clear that a localised solution is
bound to have at least one extremum in its profile. This implies
that the first derivative of $\s(\xi)$ must vanish at the extremum:
\be \label{d1} \frac{2(B-AD) f(\xi)}{(1+D {f(\xi)}^2)^2}
f^{\prime}(\xi)=0.\ee
 Since $AD{\neq}B$, either $f$ or $f^{\prime}$ or both
 must be zero. In our case $f(\xi)=sech(\xi)$,
 whose first derivative vanishes only
 at origin. This means we have an extremum at origin.
The second derivative at origin is,

\be \label{d2} \s^{\prime\prime}=\frac {2(AD-B)}{(1+D)^2},\ee

\nn which resolves the maximum and minimum. It should be noted that
$\s''$ is singular for $D=-1$. For the non-singular case, we see
that there is a clear distinction of two regimes of solutions: One
for which $AD>B$, where $\s^{\prime\prime}$ is positive; this
corresponds to {\em{minimum}}. In the second case, where $AD<B$ is
negative, we have a {\em{maximum}}. The latter, corresponds to a
{\em{bright soliton}}, whereas the former corresponds to a {\em{dark
soliton}} in the propagating media. This clearly suggests, that both
types of solitons exist in this dynamical system.

\begin{figure}
\begin{center}
\epsfxsize=4in {\epsfbox{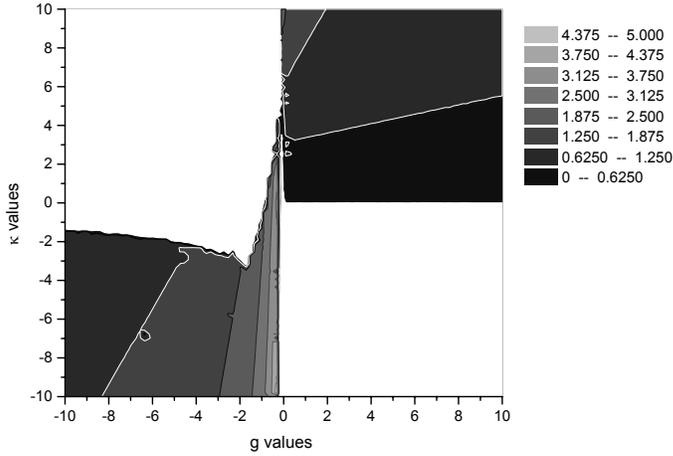}} \caption{\label{figure1} Contour
plot showing the allowed background field amplitude value $A$, for
which the localised solitary wave solution exist, here $\e=0.1$.}
\end{center}
\end{figure}

\begin{figure}
\begin{center}
\epsfxsize=4in {\epsfbox{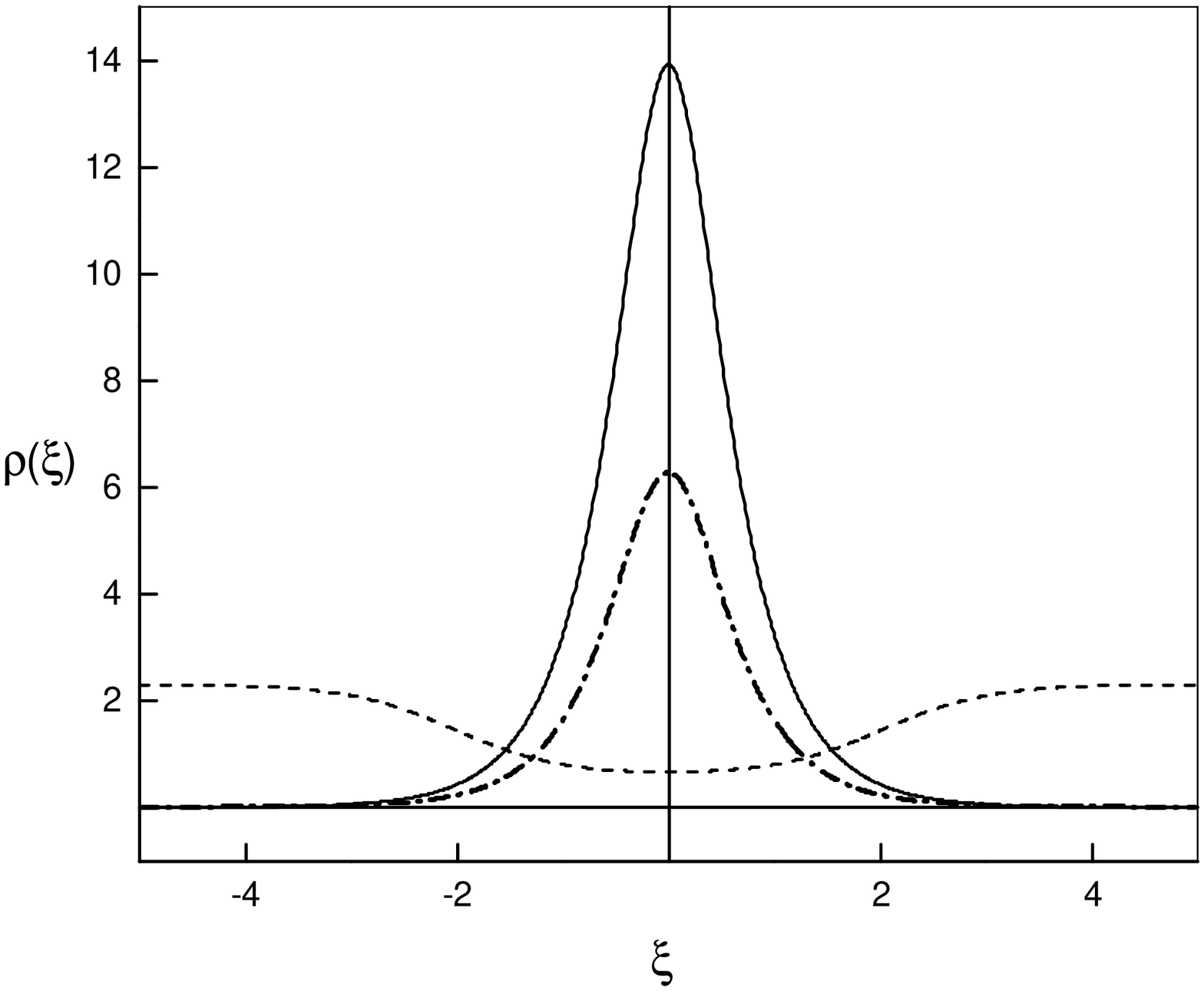}}
 \caption{\label{f2}Density
profiles of some solitons of driven NLSE.\\
i) Dark soliton (dashed), with $g=0.7$, $\kappa =-0.7$ and $\e =-1.4$;\\
ii) bright soliton (solid), with $g=-0.5$, $\kappa =0.1$ and $\e =-0.4$;\\
iii) bright soliton (dash-dotted), with $g=-0.4$, $\kappa=0.1$ and
$\e=-0.4$.}
\end{center}
\end{figure}

In these rational solutions, parameter $A$ decides the strength of
the background, in which these solutions propagate. It is
interesting to note that for the localised solution, $A=0$ is not
permitted, since this leads to the absence of the source.
Considering $\s$ to be positive semi-definite one finds further
constraints:(i)$A$ should be positive since negative background is
not meaningful; (ii)$A>-|B|$; (iii)$D$ should be greater than $-1$
for non-singularity. The parameters satisfying these conditions are
taken to be physically meaningful. The parameter conditions leading
to singular solutions are dealt separately. Figure (\ref{figure1})
illustrates various ranges of the values of the strength of the
nonlinearity $g$, and strength of the source $\k$, for solitary wave
solutions. Figure (\ref{f2}) shows the solutions for some mentioned
values of the parameters.

In case of driven HNLSE, we see that $v=0$, when the equation, \be
\label{vzero} 8 {\bar{\e}}^2 {\tilde{A}}^2 k^3 - 8 {\bar{\e}}
{\tilde{A}} k^2 + 2 k - {\bar{\e}} {\tilde{A}}({w}+\mu)=0,\ee is
satisfied. In general, the above cubic equation can have
 one real, non-zero value of $k$, for appropriate
equation parameters. Hence, we see that driven HNLSE has
non-propagating solutions even when $k{\neq}0$, which is different
from driven NLSE, since in the latter case $v=2k$.

Considering the case $k=0$, we see that, equation (\ref{realhnlse})
and equation (\ref{realeq}) have same coefficients and hence the
same solutions. However, unlike the driven NLSE for this case one
can have a propagating envelope with velocity \be \label{v} v = -
\bar{\e} \e \tilde{A}. \ee \nn
 Static solutions are obtained, when any one of
$\tilde{A}$, $\bar{\e}$ and $\e$ vanish. It is interesting to note
that when $\e$ vanishes, as we shall show later, $\a$ becomes a free
parameter, hence we get a static soliton of arbitrary size. The case
when $\bar{\e}$ vanishes, driven HNLSE simply reduces to driven
NLSE, for which $k=0$ implies $v=0$. Considering the case when the
third order dispersion parameter $\tilde{A}$ vanishes, one finds
that  for the non trivial case $\tilde{B} : \tilde{C}$ $=$ $-2 : 1$.

 We now consider the case when $\e=0$ for which
 $\omega = k^2 - \mu$; in this case equations (\ref{condi1}) to
(\ref{condi4}) yield,

 \be q_{1}(\alpha^{4})
+r_{1}\alpha^{2} + c=0.\ee
Here, $q_{1}=48A^{5} g^{2}-48 A^{2}g
\kappa$, $r_{1}=12A^{7}g^{3}-60A^{4}g{2}\kappa +48 A g\kappa^{2}$
and
 $c=-18 A^{6}g^{3}\kappa +9 A^{3}g^{2}\kappa^{2}+9g\kappa^{3}$.
Very interestingly, all of these coefficients vanish in the view of
equation (\ref{condi1}), leaving $\alpha$ as a free parameter. So,
the width of the soliton, in this case, is independent of the
parameter values, which means that the solitons can have arbitrary
size for the given values of $g$ and $\kappa$. Similar is the case
for driven HNLSE, when $\bar{\e}$ vanishes from equation
(\ref{realhnlse}).

As noted earlier, equation (\ref{realeq}) has singular solutions,
for $D=-1$. One can show that they satisfy the relation

\be g(6 {\a}^2 -3 \e) A^2 + \e (2 {\a}^2 + \e ) + 9 g \k A - 8
{\a}^4 =0. \ee

These solutions possess one singularity in their profile, which
physically corresponds to very large field intensity due to self
focussing \cite{msg,gaeta}. When $D<-1$, singular solitons exists,
with singularites existing at two different locations, between which
$\s$ becomes negative.

\section{NLSE with a generalized source}

Instead of the source term of the type we have considered
previously, one can have a more general one where,

\be \label{nlse2} i\frac{\partial \psi}{\partial t} +
\frac{\partial^{2} \psi}{\partial x^2} +g \mid \psi \mid^2 \psi +\mu
\psi=\kappa e^{i(\chi(\xi) - \omega t )}. \ee

\nn Here $\chi(\xi)$ is some function. We consider the ansatz,

 \be \label{ansatz2} \psi(x,t) =  \rho(\xi,t) e^{i(\chi(\xi) - \omega t )},\ee

\nn where $\xi=\a (x-vt)$. Equation (\ref{ansatz2}) when substituted
in equation (\ref{nlse2}) gives a complex equation in $\rho$ and
$\chi$. Equating the imaginary part to zero, one gets,

\be \label{imag2} \chi^{\prime} = \frac {v}{2\alpha} +
\frac{c}{\rho^{2}}.\ee

\nn One can clearly see that the above equation suggests
phase-amplitude coupling for $c{\neq}0$. We see from relation
(\ref{imag2}) an interesting case of the phase singularity of the
soliton arising through phase-amplitude coupling, when $\r$
vanishes.

\nn Substituting this relation in the real part of equation
(\ref{nlse2}), we get a non linear differential equation of the
form,

\be \label{realeq2} \alpha^2 \rho^{\prime\prime} +g\rho^3+\e \rho =
\kappa + \frac {\alpha^{2} c^{2}} {\rho^{3}},\ee

\nn where $\e=(\omega +\mu +\frac{v^2}{4})$. The above equation can
be connected with the equation governing Jacobi elliptic functions
via a fractional transformation as,

 \be \label{FT1}   \rho (\xi) =  \frac {A + B f(\xi|m)^{\d}} {1+
D f(\xi|m)^{\d}},\ee

\nn where $\d=2$ is again the maximum allowed value. This gives the
 consistency conditions:

\bea
\label{r1} gA^6 +\e A^4 -(2\alpha^2 (AD-B)(1-m)+\kappa)A^3 =\alpha^2 c^2 \eea 

\bea \label{r2} 6A^5 B g + 4 \e A^3 B + 2 \e A^4 D - 3 \kappa A^2 B
- 3 \kappa A^3 D+
            6 \alpha^2 (B-AD)(1-m)A^2 B + \nonumber \\6 \alpha^2 (AD-B)(1-m)A^3 D
               +4\alpha^2 (2m-1)A^3 (B-AD)=6\alpha^2 c^2 D  \eea
\bea \label{r3} 6\alpha^2 (B-AD)(1-m)A B^2 + 18
\alpha^2(AD-B)(1-m) A^2 B D+\nonumber \\ 12 \alpha^2
(B-AD)(2m-1)A^2 B + 4\alpha^2 (AD-B)(2m-1)A^3 D +\nonumber \\6
\alpha^2 (AD-B)m A^3 + 15 g A^4 B^2 + 6\e A^2 B^2 + 8\e A^3 BD
+\e A^4 D^2 \nonumber \\-3\kappa A B^2 - 9\kappa A^2 B D - 3\kappa A^3 D^2 = 15\alpha^2 c^2 D^2 \eea

\bea \label{r4} 20 g A^3 B^3 + 4\e A B^3 + 12\e A^2 B^2 D + 4\e
A^3 B D^2 \nonumber \\ -\kappa B^3-9\kappa A B^2 D -9\kappa A^2 B
D^2 - \kappa A^3 D^3 + \nonumber \\18\alpha^2 A B^2 D
(AD-B)(1-m)-2\alpha^2 B^3 (AD-B)(1-m) - \nonumber \\12\alpha^2 A
B^2 (AD-B)(2m-1)-2\alpha^2 m A^3 D (AD-B) \nonumber \\
+12\alpha^2 A^2 B D (AD-B)(2m-1)=20 \alpha^2 c^2 D^3 \eea 

\bea \label{r5} 15 g A^2 B^4 +\e B^4 +8 \e A B^3 D + 6\e A^2 B^2
D^2\nonumber \\ -3\kappa B^3 D - 9\kappa A B^2 D^2-3\kappa A^2 B D^3
 + 6\alpha^2 B^3 D (AD-B)(1-m)\nonumber \\-4\alpha^2 B^3
(AD-B)(2m-1)+12 \alpha^2 A B^2 D
(AD-B)(2m-1) + \nonumber \\18\alpha^2 m A B^2 (AD-B)-6\alpha^2 m A^2 BD (AD-B)
=15 \alpha^2 c^2 D^4 \eea 

\bea \label{r6} 6 g A B^5 + 2\e B^4 D + 4\e A B^3 D^2 -3 \kappa A
B^2 D^3 \nonumber \\-3\kappa B^3 D^2 + 4 \alpha^2 (B-AD)(1-2m)B^3
D \nonumber\\+ 6 \alpha^2 m (B-AD)A B^2 D + 6\alpha^2 m (AD-B)B^3
        =6\alpha^2 c^2 D^5 \eea 
\bea \label{r7} 2\alpha^2 (B-AD)m B^3 D + gB^6 +\e B^4 D^2 -\kappa
B^3 D^3 =\alpha^2 c^2 D^6 \eea

Since, there in total are seven simultaneous relations to fix three
independent parameters, we can see that these are constrained
solutions. One can also see that the phase-amplitude coupling
imposes constraints on the solution space, without changing the
profile of the solutions. From the above relations, one can see that
$B=0$ needs $c=0$ or $D=0$. The former case has already been studied
here, whereas the latter case results in a constant background
solution. Here, $A{\neq}0$ requires $c{\neq}0$, hence the solutions
always exist in a constant background. This should be contrasted
with the case when $c=0$; for which cnoidal wave solutions are
possible with $A=0$. Similarly, in $m=1$ case localised solutions
with $B=0$ are allowed in the previous case, which as seen above are
not found in present case. Hence, only rational solutions are
possible. Moreover, we see that (\ref{r1}) is a sixth order
polynomial, in contrast to the previous relations, this does not
have analytically tractable roots leaving numerical analysis as the
only tool to analyse the structure of solution space.

In conclusion, a number of interesting features have emerged from
analysis of the exact solutions of the driven NLSE and HNLSE. Dark
and bright solitons can exist in attractive and repulsive non-linear
regimes, a feature very different for NLSE; the presence of the
external source makes this possible. In a wide range of parameters,
the driven HNLSE and driven NLSE have similar solutions, albeit with
different sizes and velocities. Static solitons are found for both
the equations. In particular, static solitons are found to have
interesting properties like arbitrary scale, under certain
parametric restrictions. They differ significantly for driven NLSE
and HNLSE cases.
 In certain specific parameter regimes, solitons of arbitrary size are found. Singular solutions are also
found for both the equations. The investigation of the situation,
where the phase is allowed to depend upon the intensity, revealed
that  the corresponding solutions are highly constrained. The study
of solitary waves having complex envelope, analogous to Bloch
solitons in condensed matter physics, is worth investigating in the
present scenario. Application of the fractional linear
transformation technique employed here to other nonlinear equations
is also of deep interest. Investigations along these lines are
currently in progress.

\vskip.5cm {\bf{References}} 
\end{document}